\documentclass[journal]{IEEEtran}

\usepackage{amsmath}
\usepackage{amssymb}
\usepackage{amsfonts}
\usepackage{graphicx}
\usepackage{epsfig}
\usepackage{subfigure}
\usepackage{psfrag}
\usepackage{cite}
\usepackage{latexsym}
\usepackage{url}
\usepackage{color}
\usepackage{multirow}
\usepackage{mathtools}
\usepackage{bm}
\usepackage{booktabs}
\usepackage{algorithm}
\usepackage{algpseudocode}
\usepackage{bm}

\usepackage{verbatim}
\usepackage{indentfirst}
\usepackage{hyperref}
\usepackage{multirow}
\usepackage{bbm}

\graphicspath{{fig/}}

\PassOptionsToPackage{bookmarks={false}}{hyperref}

\IEEEoverridecommandlockouts

\allowdisplaybreaks[4] 

\begin{document}
\title{Empowering Intelligent Low-altitude Economy with Large AI Model Deployment}
\author{
Zhonghao Lyu, \emph{Member, IEEE},   Yulan Gao, \emph{Member, IEEE}, Junting Chen, \emph{Member, IEEE},\\ Hongyang Du, \emph{Member, IEEE},  Jie Xu, \emph{Fellow, IEEE},  Kaibin Huang, \emph{Fellow, IEEE}, \\ and Dong In Kim, \emph{Life Fellow, IEEE}
\thanks{Z. Lyu, J. Chen, and J. Xu are with the School of Science and Engineering, the Shenzhen Future Network of Intelligence Institute, and the Guangdong Provincial Key Laboratory of Future Networks of Intelligence, The Chinese University of Hong Kong (Shenzhen), China  (e-mail: zhonghaolyu@link.cuhk.edu.cn, juntingc@cuhk.edu.cn, xujie@cuhk.edu.cn).}
\thanks{Y. Gao is with the Department of Information Science and Engineering, KTH Royal Institute of Technology, Sweden (e-mail: yulang@kth.se).}
\thanks{H. Du and K. Huang are with the Department of Electrical and Electronic Engineering, University of Hong Kong, Hong Kong SAR, China (e-mail: duhy@eee.hku.hk, huangkb@hku.hk).}
\thanks{D. I. Kim is with the Department of Electrical and Computer Engineering,
Sungkyunkwan University, South Korea (email: dongin@skku.edu).}
}

\maketitle

\begin{abstract}
Low-altitude economy (LAE) represents an emerging economic paradigm that redefines commercial and social aerial activities. Large artificial intelligence models (LAIMs) offer transformative potential to further enhance the intelligence of LAE services. However, deploying LAIMs in  LAE poses several challenges, including the significant gap between their computational/storage demands and the limited onboard resources of LAE entities, the mismatch between lab-trained LAIMs and dynamic physical environments, and the inefficiencies of traditional decoupled designs for sensing, communication, and computation.
To address these issues, we first propose a hierarchical system architecture tailored for LAIM deployment and present representative LAE application scenarios. Next, we explore key enabling techniques that facilitate the mutual co-evolution of LAIMs and low-altitude systems, and introduce a task-oriented execution pipeline for scalable and adaptive service delivery. Then, the proposed framework is validated through real-world case studies. Finally, we outline open challenges to inspire future research.
\end{abstract}

\section{Introduction}
The low-altitude economy (LAE) is rapidly emerging as a critical engine of global industrial innovation and economic growth. As a foundational element of next-generation digital infrastructure, LAE is enabling transformative services in logistics, transportation, and public safety. For instance, China’s LAE market is expected to exceed 1 trillion RMB by 2026 with a 33.8$\%$ annual growth rate. Germany has investigated over 13 billion USD to air mobility since 2019, while Amazon’s Prime Air MK30 drones have been designed for corss-country package delivery in 2024 \cite{YYang}. These global developments highlight the urgent need for scalable, intelligent, and secure low-altitude systems.

In parallel, the rise of large artificial intelligence models (LAIMs) has redefined the capabilities of machine perception, reasoning, and decision-making beyond traditional AI systems, offering a powerful foundation for intelligent LAE services. Their strong generalization capabilities and cross-domain adaptability make them well-suited for various tasks, such as aerial monitoring \cite{JWang}, robust communication \cite{SJavaid, CZhao}, and multi-agent coordination \cite{LZhou}. 

Despite their promise, deploying LAIMs in LAE introduces several  challenges: 

1) {\it Resource-aware deployment:} Generally, the massive model size and high computation intensity far  exceed the onboard battery, storage, and processing capabilities of intelligent aerial agents (IAAs) and ground nodes. This makes direct deployment of full-scale LAIMs infeasible. 
To tackle this issue, recent efforts have explored scalable LAIM deployment strategies, including efficient pre-training and fine-tuning \cite{ZLyu1}, joint task-offloading and resource management \cite{YGao}, and light-weight model inference \cite{ZLyu2}.
Moreover, full-scale LAIMs can be hosted on the edge/cloud servers with sufficient computing power, while IAAs interact with them for prediction and inference via prompt engineering \cite{HLi}.

2) {\it Environment-aware adaptation:} Low-altitude system design must adapt to highly complex and dynamic real-world environments. This includes handling three-dimensional (3D) airspace geometry, operation constraints, electromagnetic interference, and seamless integration with terrestrial and satellite networks. Conventional offline designs built on idealized simulation environments fail to capture such dynamics, necessitating environment-aware IAA deployment and real-time resource management \cite{YZeng}. Furthermore, the performance of LAIMs pre-trained on general-purpose datasets may degrade, when faced with mission-specific tasks and evolving real-world scenarios. To ensure effectiveness, LAIMs must be continuously adapted through feedback-driven refinement informed by real-world interactions and task-specific data.

3) \emph{Co-designing of  sensing, communication, and computation (SCC):} To meet heterogeneous quality-of-service (QoS) requirements in LAE, it is essential to co-design SCC in a task-oriented paradigm. Traditional decoupled designs often lead to resource inefficiencies, as the three functionalities are often closely coupled and may compete for constrained resources in low-altitude systems \cite{DWen}. A task-oriented integrated SCC (ISCC) framework could ensure task-relevant sensing, semantic-aware transmission, and adaptive computation tailored to specific LAIM tasks.

\begin{figure*}[h]
	\centering
	 \epsfxsize=1\linewidth
	\includegraphics[width=18cm]{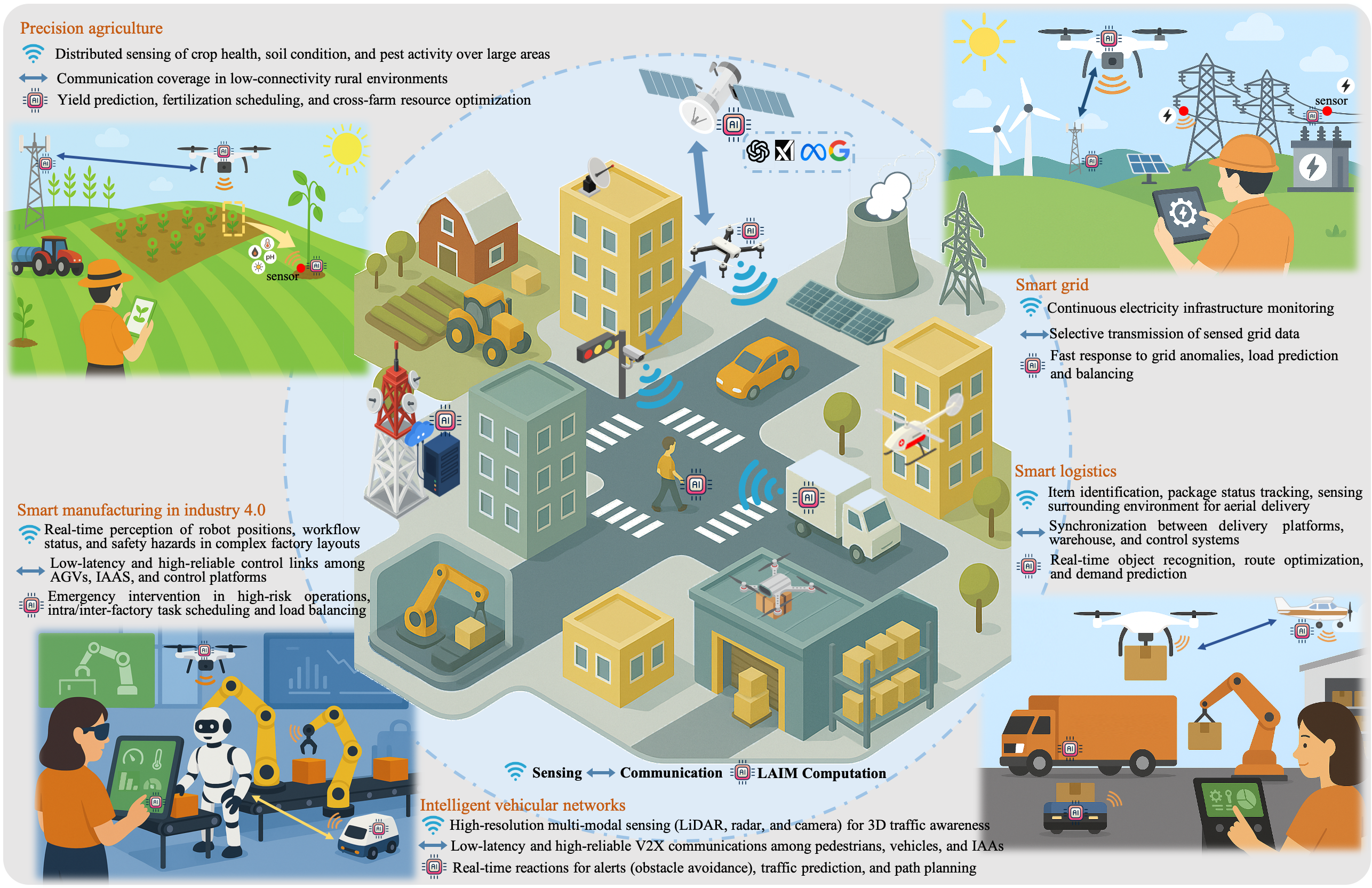}
		\vspace{-7pt}
	\caption{\label{model} Architecture and application scenarios of LAIM-empowered intelligent LAE services.}
	\vspace{-10pt}
\end{figure*}

These challenges call for unified system architectures, environment-aware system design methodologies, and real-world prototypes to enable efficient deployment of LAIMs to empower intelligent LAE services, which are still underexplored in existing literature. Motivated by this gap, this paper investigates the integration of LAIMs into low-altitude systems. The main contributions are summarized as follows,
\begin{itemize}
\item First, we propose a hierarchical system architecture for deploying LAIMs in low-altitude systems, and classify representative LAE application scenarios. Also, open issues are outlined to motivate future research.
\item Second, we explore the mutual benefits between LAIMs and LAE, including key enabling techniques and real-world interaction loops that continuously refine LAIM performance. Moreover, we propose a task-oriented execution pipeline for LAIM-empowered LAE.
\item Finally, we present real-world case studies, i.e., LAIM-guided uncrewed aerial vehicle (UAV) placement and resource-aware LAIM deployment across air-ground platforms. These studies are grounded in real-world data and testbeds, demonstrating the feasibility of our framework under practical low-altitude environments.
\end{itemize}

\section{System Architecture and Application Scenarios of LAIM-empowered LAE}

LAIM-empowered LAE requires requires a fundamental shift in architectural design for intelligent service delivery. Unlike traditional mobile edge computing (MEC) systems that rely on static terrestrial infrastructure, LAE necessitates an aerial-native and mission-adaptive system architecture capable of supporting distributed AI execution across dynamic 3D network topologies with fast-moving IAAs. To address these unique demands, we propose a three-tier hierarchical framework, as shown in Fig.~\ref{model}, consisting of a ground layer, an aerial layer, and a cloud layer, which enables scalable and adaptive LAIM-empowered intelligence across LAE services.

\subsection{Low-altitude System Architecture with LAIM Integration}

The ground layer comprises heterogeneous terrestrial nodes, such as sensors, mobile user equipments, base stations (BSs), and edge servers, which are responsible for ubiquitous data collection and preliminary processing. While sensors and mobile user equipments are typically constrained by limited battery, memory, and processing resources, making them suitable only for lightweight LAIM components like ultra-compact encoders for early-stage feature extraction, BSs and edge servers offer stronger processing and communication capabilities. This hierarchical architecture enables the transmission of semantically compressed embeddings or task-relevant raw data to higher-tier nodes, reducing bandwidth consumption while preserving task-critical information.

The aerial layer consists of IAAs, such as UAVs, which serve as multifunctional entities integrating sensing, computing, communication, and service execution. 
Unlike traditional ground infrastructures, IAAs offer a unique aerial perspective, enabling the capture of comprehensive 3D environmental views and situational awareness that is difficult to obtain from the ground. Their high mobility allows for rapid repositioning and agile adaptation, supporting flexible network topology reconfiguration, on-demand coverage extension, and load balancing in response to dynamic service demands or environmental changes.
For instance, in logistics or disaster response scenarios,  IAAs act as aerial users to simultaneously perceive the environment, process data in real time, and maintain reliable low-latency communications with centralized controllers. 
When deployed as aerial BSs, they extend network coverage in hotspots or emergency zones, improving connectivity and service continuity. Each IAA is embedded with segmented LAIM modules and edge computing capabilities to execute latency-sensitive tasks locally, while offloading compute-intensive tasks to the cloud for collaborative processing. 
Moreover, in cooperative tasks, IAAs form flying ad hoc networks through  dynamic routing, adaptive beamforming, and mesh reconfiguration, offering robust inter-agent connectivity and resilience  under high mobility, interference, or partial network failure, far beyond what static ground infrastructures can achieve.

The cloud layer consists of centralized infrastructure, including terrestrial BSs, cloud data centers, and low-orbit satellites. It acts as the global coordinator for LAIM-empowered LAE services, supporting distributed model training, inference, and cross-layer resource management. It typically hosts various full-scale LAIMs and partitions their execution across different layers based on environment-aware and task-specific QoS requirements, such as task urgency, channel conditions, and resource availability. This hierarchical orchestration enables flexible LAIM deployment for large-scale LAE applications in real-world.

\subsection{Typical Application Scenarios}
Typical LAE applications can be broadly categorized into three classes based on their dominant functional demands, i.e., communication-centric, sensing-centric, and computation-centric scenarios.

\subsubsection{Communication-centric Applications} 
The primary challenge is to achieve intelligent, reliable, and low-latency communication in dynamic and resource-limited low-altitude environments. Unlike traditional UAV-assisted communication networks, LAIM-enabled IAAs not only serve as aerial BSs, extending wireless coverage in disaster zones, temporary hotspots, or infrastructure-limited rural regions, but also perform real-time scheduling and dynamic spectrum access. Moreover, IAAs can form self-organized mesh networks beyond static relaying, to support robust multi-hop communications, ensuring continuity between mobile agents and disconnected infrastructure segments.
Additionally, IAAs leverage multi-sensor fusion for precise wireless channel measurement and modeling, enabling predictive channel estimation, 3D beamforming, and interference mitigation, thereby providing a resilient communication backbone for LAE networks.


\subsubsection{Sensing-centric Applications}
Sensing-centric scenarios focus on air-ground collaborative and high-resolution environmental awareness, achieving by the enhanced perception capabilities of IAAs equipped with multi-modal sensors, such as LiDAR, cameras, and millimeter-wave radar.
Unlike traditional UAVs, LAIM-empowered IAAs are capable of dynamic sensing adaptation,  which autonomously modify sensing modes and trajectories in response to real-time environmental feedback. Representative use cases include precision agriculture, urban pollution assessment, and disaster response, where periodic scanning, wide-area coverage, and multi-angle imaging are essential to detect crop stress, evaluate air quality, or rescue survivors in post-disaster areas. Another critical application is obstacle detection and avoidance during aerial navigation for transportation and delivery missions, particularly in low-visibility or rapidly changing environments, where IAAs  continuously observe their surroundings to identify potential hazards such as trees, power lines, or other flying objects, enabling timely and safe trajectory adjustments.

\subsubsection{ Computation-centric Applications}
Computation-centric applications position LAE entities (i.e., ground, aerial, and cloud nodes) as distributed computation infrastructures, supporting a wide range of AI-driven tasks. Specifically, for delay-sensitive tasks, such as anomaly detection in smart grids or real-time UAV path replanning, IAAs act as mobile edge computing nodes, by processing data close to the source for fast computation responses. Beyond real-time services, LAE also supports computation-intensive applications that require large-scale data analysis, predictive reasoning, and global optimization. Examples include low-altitude channel and environment map reconstruction, traffic forecasting, and airspace coordination through system-level optimization. Nevertheless, continuous LAIM fine-tuning via real-world interactions is necessary for narrowing the performance gap between lab training and field deployment.

To summarize, Fig. \ref{model} presents representative use cases of LAIM-empowered LAE, including smart manufacturing, smart grid, intelligent vehicular networks, smart logistics, and precision agriculture, while highlighting the corresponding sensing, communication, and computation functionalities involved in each scenario.

\section{Mutual Benefits of Deploying LAIMs in LAE}
This section discusses the mutual benefits of integrating LAIMs into LAE, and presents a task-oriented execution pipeline to facilitate this integration.

\begin{figure*}[h]
	\centering
	 \epsfxsize=1\linewidth
	\includegraphics[width=14cm]{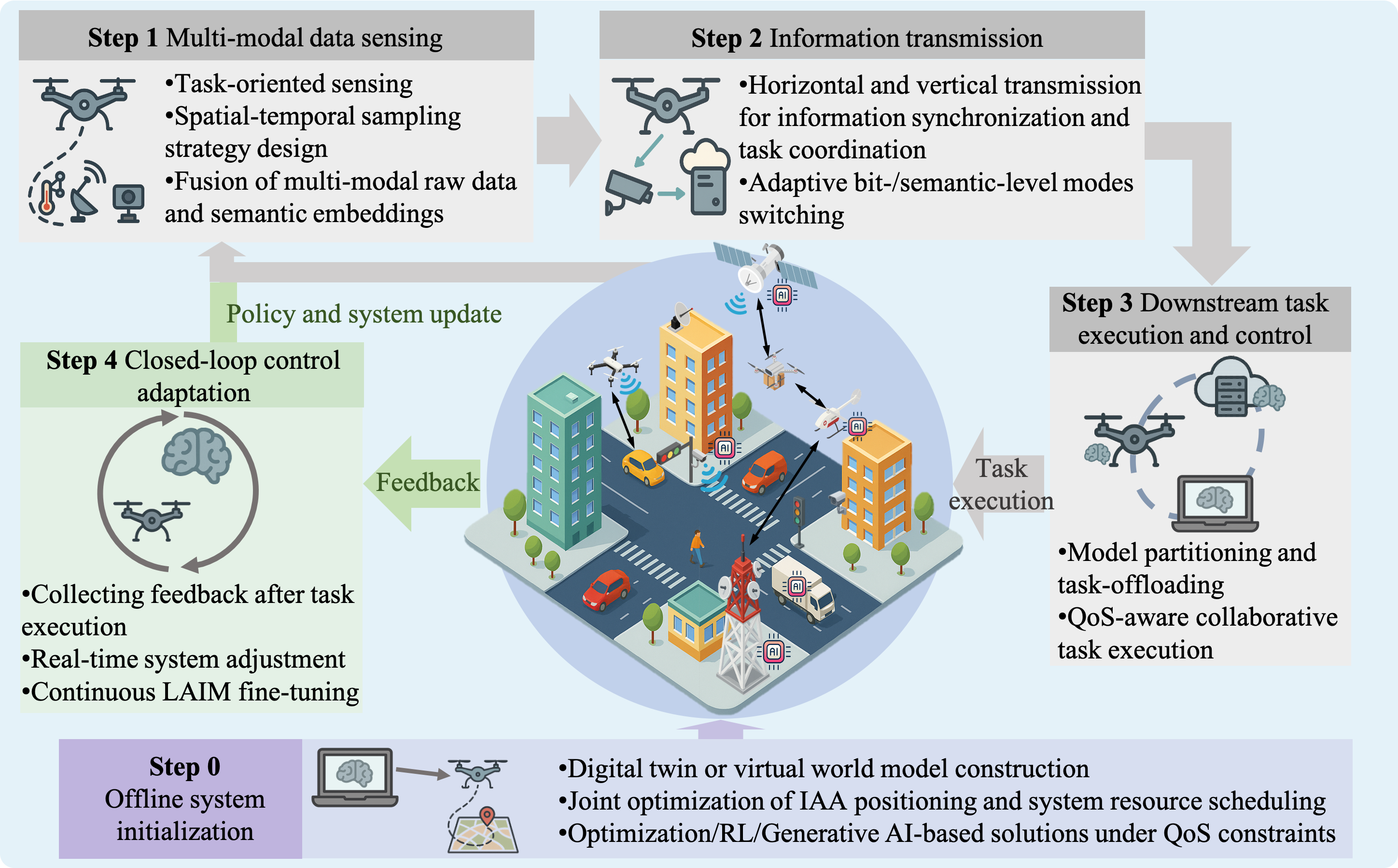}
		\vspace{-7pt}
	\caption{\label{framework} Execution pipeline for LAIM-empowered LAE services.}
	\vspace{-10pt}
\end{figure*}

\subsection{Key Enabling Techniques for LAIM-empowered LAE}
Deploying LAIMs in dynamic and resource-constrained low-altitude environments requires tight integration of SCC functionalities to ensure real-time, mission-adaptive intelligence. We identify three core enabling techniques that jointly empower LAIM deployment: integrated sensing and communication (ISAC), integrated communication and computation (ICC), and ISCC.

\subsubsection{ISAC for LAIM-empowered LAE}
In low-altitude systems, multi-modal sensors are widely deployed on both IAAs and ground nodes. ISAC frameworks enable these entities to use shared spectrum and hardware resources for both environmental sensing and data transmission by leveraging dual-purpose waveform design \cite{ZLyu3}. This design paradigm is especially beneficial in low-altitude scenarios where spectrum availability and hardware capacity are constrained. Beyond resource reuse, ISAC enables environment-aware channel modeling and real-time channel state information (CSI) acquisition by jointly estimating the channel parameters and perceiving the physical environment \cite{WLiu}. Moreover, network-level ISAC further enables adaptive 3D beamforming, allowing IAAs and edge nodes to sustain high-quality communication links and perform high-precision sensing collaboratively, even in highly dynamic or low-visibility environments.

\subsubsection{ICC for LAIM-empowered LAE}
ICC focuses on the co-design of communication and computation to ensure efficient transmission and processing of task-relevant information with reduced latency. Two foundational enablers of ICC in LAIM-empowered LAE are semantic communications and over-the-air computation (AirComp). 
Semantic communications aim to transmit only compressed task-relevant features or intermediate inference results instead of raw data. 
For example, \cite{YCheng1} proposed an interest-based image segment transmission method, which can help to extract only the relevant part of UAV-sensed images to reduce bandwidth consumption and latency.
AirComp exploits the superposition property of wireless multiple-access channels to enable simultaneous communication and functional aggregation of distributed data. 
This is particularly advantageous for collaborative LAE tasks needing efficient information aggregation, such as UAV swarming, distributed sensing, and consensus building. 
Together, these techniques enable a compute-while-communicate paradigm, alleviating communication bottlenecks and enhancing the responsiveness of latency-sensitive LAIM-empowered services in  resource-constrained low-altitude environments.

\subsubsection{ISCC for LAIM-empowered LAE}
ISCC represents an advanced convergence of ISAC and ICC, enabling the task-driven co-optimization of SCC resources. Traditionally designed in isolation, these functions are inherently coupled and exhibit complex relationships in LAIM-empowered low-altitude systems, from collaboration to competition over limited resources. The overall service quality depends on their joint effectiveness, but their contributions to task performance are uneven. Therefore, their individual optimization, e.g., maximizing sensing resolution or transmission rate, can lead to degraded task performance and inefficiencies.

To address it, a task-oriented ISCC design paradigm is essential, which co-optimizes SCC for mission-level goals such as LAIM training/inference accuracy, latency, energy efficiency, and reliable primary and secondary control. In this paradigm, sensing prioritizes acquiring task-relevant information, communication focuses on transmitting semantic features or intermediate results, and computation is adaptively offloaded across heterogeneous entities aligned with task-specific QoS demands \cite{ZLyu4}. Additionally, jointly optimizing IAA positioning with SCC resource allocation further enhances system flexibility and supports adaptation to the dynamic low-altitude environments.
\begin{itemize}
    \item {\bf ISCC for distributed LAIM training in LAE:} ISCC provides the architectural backbone for LAIM-based services in LAE, supporting continual model refinement driven by real-world feedback. Emerging techniques, such as parameter-efficient fine-tuning (e.g., low-rank adaptation (LoRA) and prompt-tuning), mixture-of-experts (MoE), and distributed learning (e.g., federated and split learning) are particularly suitable for life-long LAIM adaptation and naturally aligned with ISCC  principles in low-altitude environments. For example, to fine-tune a pre-trained LAIM for traffic monitoring, split learning can be used to distribute model segments and training tasks across ground devices, IAAs, and the cloud, leveraging heterogeneous computation resources. Under this setup, only gradient/model information needs to be exchanged, reducing communication overhead and enhancing privacy compared to transmitting raw training data. LoRA can accelerate this fine-tuning process, while federated learning enables semantic heads to generalize across multi-modal and multi-task settings via periodic model aggregation. Throughout this process, a task-aware ISCC framework coordinates model partitioning, node selection, resource allocation, and IAA positioning to accelerate the convergence speed.
    \item {\bf ISCC for efficient LAIM inference in LAE:} ISCC also supports efficient inference by enabling flexible partitioning and execution of LAIMs across ground, aerial, and cloud nodes based on real-time QoS requirements, which is commonly referred to as split or co-inference. 
    Enhanced by expert-path routing, model components such as attention heads or decision layers, can be selectively executed depending on the capabilities of each node. Moreover, to enable lightweight deployment for fast inference, various model compression techniques, such as pruning, quantization, sparse activation, and knowledge distillation are applied.
    However, these must be guided by QoS-aware strategies and theoretical performance guarantees to ensure a balance among inference accuracy, latency, and energy consumption. For example, high-reasoning-demand tasks (e.g., multi-view 3D scene reconstruction) may require retaining more model capacity, even at higher energy and delay costs.
    ISCC coordinates model compression, task offloading, IAA placement, and resource allocation, enabling resource-efficient LAIM inference tailored to evolving service demands in dynamic low-altitude environments.
\end{itemize}

\subsection{LAE for Continual LAIM Evolution}
While LAIMs serve as powerful enablers of intelligent LAE services, the relationship between them is inherently bidirectional. LAE is not merely a consumer of LAIMs, it could also help to advance LAIMs. 
Unlike static datasets or simulation environments, low-altitude systems continuously generate vast volumes of real-world, multi-modal, spatio-temporal data under diverse physical constraints, enabling task-oriented fine-tuning of pre-trained LAIMs.

\subsubsection{Real-World Data for Robust Representation Learning}
The multi-modal and spatially distributed sensing data produced in low-altitude environments often contain rich contextual correlations, varying noise patterns, and real-world outliers. These characteristics are critical for learning robust and generalizable representations. Continuous exposure to such diverse data fosters improved feature extraction, contextual adaptation, and multi-modal data fusion, allowing LAIMs to refine digital twin-based world models and semantic knowledge bases.

\begin{figure*}[h]
	\centering
	 \epsfxsize=1\linewidth
	\includegraphics[width=18cm]{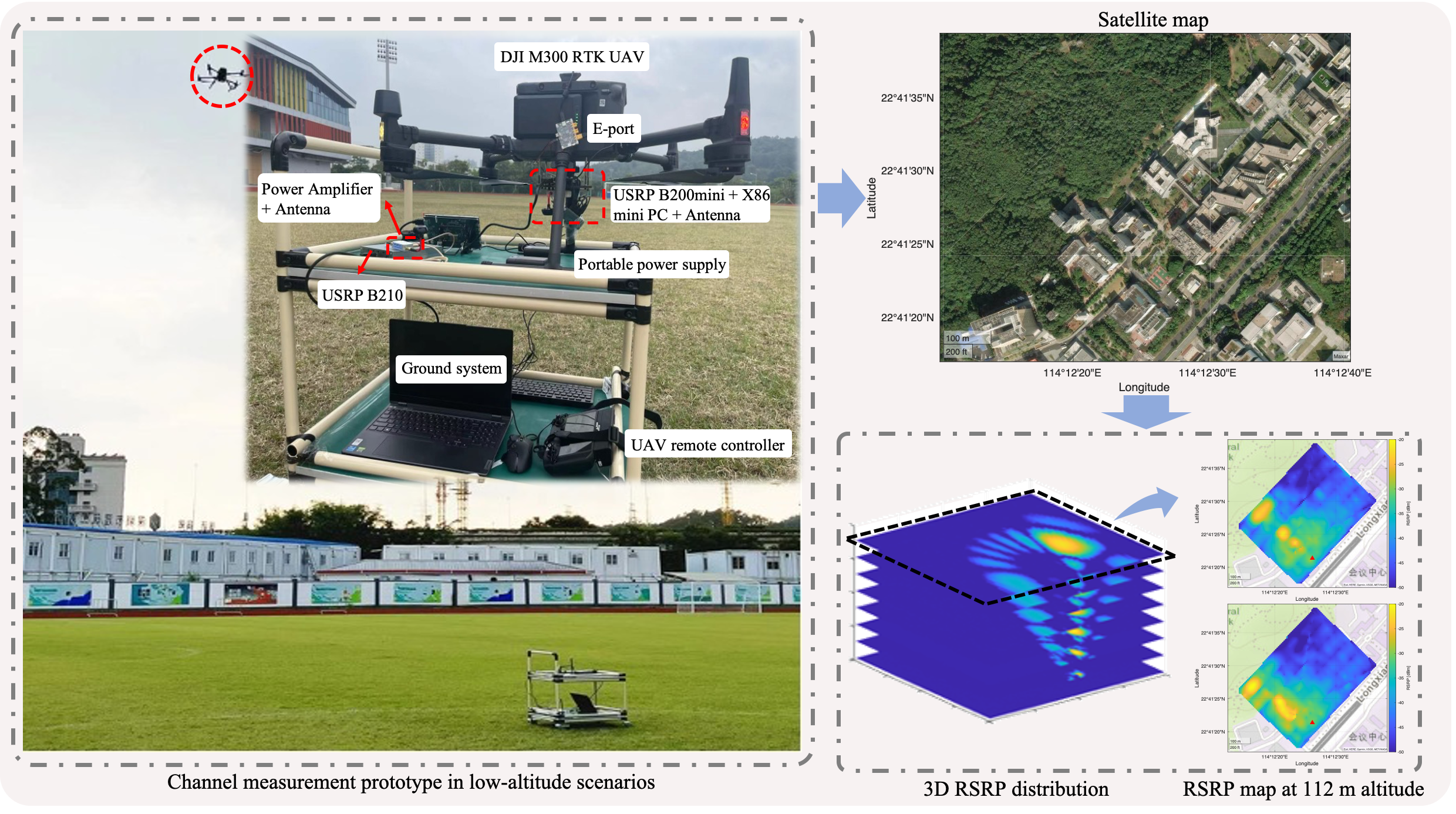}
	\vspace{-6pt}
	\caption{\label{prototype} Real-world implementation with the channel measurement prototype in low-altitude scenarios.}
	\vspace{-6pt}
\end{figure*}

\subsubsection{Reinforcement Learning (RL) in Real-world Loops}
LAE offers unique training environments for RL-type LAIMs, where intelligent agents interact directly with the physical world, execute tasks, observe feedback, and iteratively improve policies. The complexity of LAE tasks introduces high-dimensional state-action spaces and diverse reward mechanisms, which are ideal for training scalable RL architectures. These systems can be further enhanced using human-in-the-loop feedback, such as expert demonstrations, to guide policy optimization. Through continual interaction with real-world environments, LAIMs can develop reusable policies that generalize to unseen conditions.

\subsubsection{Multi-Task LAIM Adaptation}
The heterogeneity of LAE tasks necessitates the deployment of multi-task LAIMs for cross-domain reasoning. A single IAA may be expected to perform tasks from visual inspection to path planning, which pose significant challenges to traditional single-task models.
To meet these demands, LAIMs are evolving toward rapid multi-task adaptation, adopting parameter-efficient designs, such as shared multi-head attention, cross-modal encoders, and task-specific output modules. Moreover, low-altitude systems inherently support  continual LAIM learning, where task failures naturally trigger on-the-fly adjustment and retraining.

\subsection{Execution Pipeline}
The execution of LAIM-empowered LAE services follows a task-oriented pipeline comprising both offline and online stages, which is summarized as follows.

\subsubsection{Offline System Initialization} 

The pipeline begins with an offline initialization phase grounded in a digital twin or virtual world model. At this stage, IAA positioning, resource allocation, cooperation strategies, and task scheduling are jointly optimized under task-oriented objectives and QoS constraints. This optimization problem can be solved via a range of techniques, including convex/non-convex optimizations, RL, and generative AI. The outcome provides a coarse-grained initial system setups. However, due to inevitable mismatches between simulation and real-world conditions, this offline plan requires subsequent refinement through online adaptation.

\subsubsection{Multi-modal Data Sensing}
Once deployed, LAE nodes begin collecting environmental data using multi-modal sensors mounted on both ground terminals and IAAs. The sensing process is task-oriented, employing spatially selective and temporally adaptive strategies, such as continuous or event-triggered sampling, based on region-of-interest characteristics and mission requirements. The collected data, ranging from raw sensing data to semantic features, form the foundation for downstream LAIM inference and control tasks.

\subsubsection{Information Transmission} 
To synchronize information and support cooperative intelligence, sensing and control signals are transmitted across the LAE networks. This involves both horizontal exchange within each layer (e.g., among IAAs or ground terminals) to build  task and environment consensus, and vertical exchange across layers (e.g., IAA-to-cloud) for collaborative task execution.  Depending on task urgency, node topology, and available resources, the system dynamically chooses between bit-level and semantic-level transmission modes to improve communication efficiency.

\subsubsection{Downstream Task Execution and Control} 
LAIM tasks are executed in a distributed and collaborative manner across LAE entities. Execution is partitioned based on task complexity and latency sensitivity. For example, lightweight semantic heads may run on ground nodes, transformer encoders on IAAs, and global decision modules on the cloud. This paradigm supports both real-time responsiveness for latency-critical operations and high-accuracy performance for computationally intensive tasks requiring centralized processing.

\subsubsection{Closed-loop Control Adaptation}
Following task execution, the system continuously monitors environmental feedback, such as task outcomes, signal quality variations, and unexpected disturbances. These feedback signals are used to update resource allocation, adjust IAA deployment strategies, and reconfigure task execution pipelines in real time. Through this closed-loop control mechanism, the system supports timely performance evaluation, strategy adjustment, and LAIM refinement under dynamic and uncertain low-altitude environments.

\section{Case Study}
We present case studies to validate the proposed framework, focusing on LAIM-guided UAV deployment and resource-aware LAIM deployment across LAE platforms.

\subsection{Real-world Implementation Details}
We conduct signal mapping experiments using a low-altitude channel measurement prototype  over the campus of the Chinese University of Hong Kong, Shenzhen. 
The BS is equipped with a $4 \times 8$ dual-polarized antenna array and periodically transmits $7$ synchronization signal block (SSB) beams. We built a USRP-based software-defined radio platform, mounted on a DJI M300 RTK UAV, to collect raw I/Q samples over a 10 MHz bandwidth at 2.6 GHz. The UAV flies at a fixed altitude of 112 meters above the BS, and maintains a speed of 10 m/s over an area of approximately $530 \times 400$ meters. The collected I/Q samples are processed following the 5G physical layer signal processing protocols, where SSBs are detected and their reference signal received power (RSRP) is estimated. 
RSRP values are sampled at 10 Hz from the PBCH-XSS RSRP metric, while RTK-based positioning ensured centimeter-level localization accuracy. 
The collected data provides a fine-grained signal quality measurements under realistic low-altitude environments as shown in Fig. \ref{prototype}.

\vspace{-10pt}
\subsection{LAIM-guided UAV Deployment}
\begin{figure*}[h]
	\centering
	 \epsfxsize=1\linewidth
	\includegraphics[width=18cm]{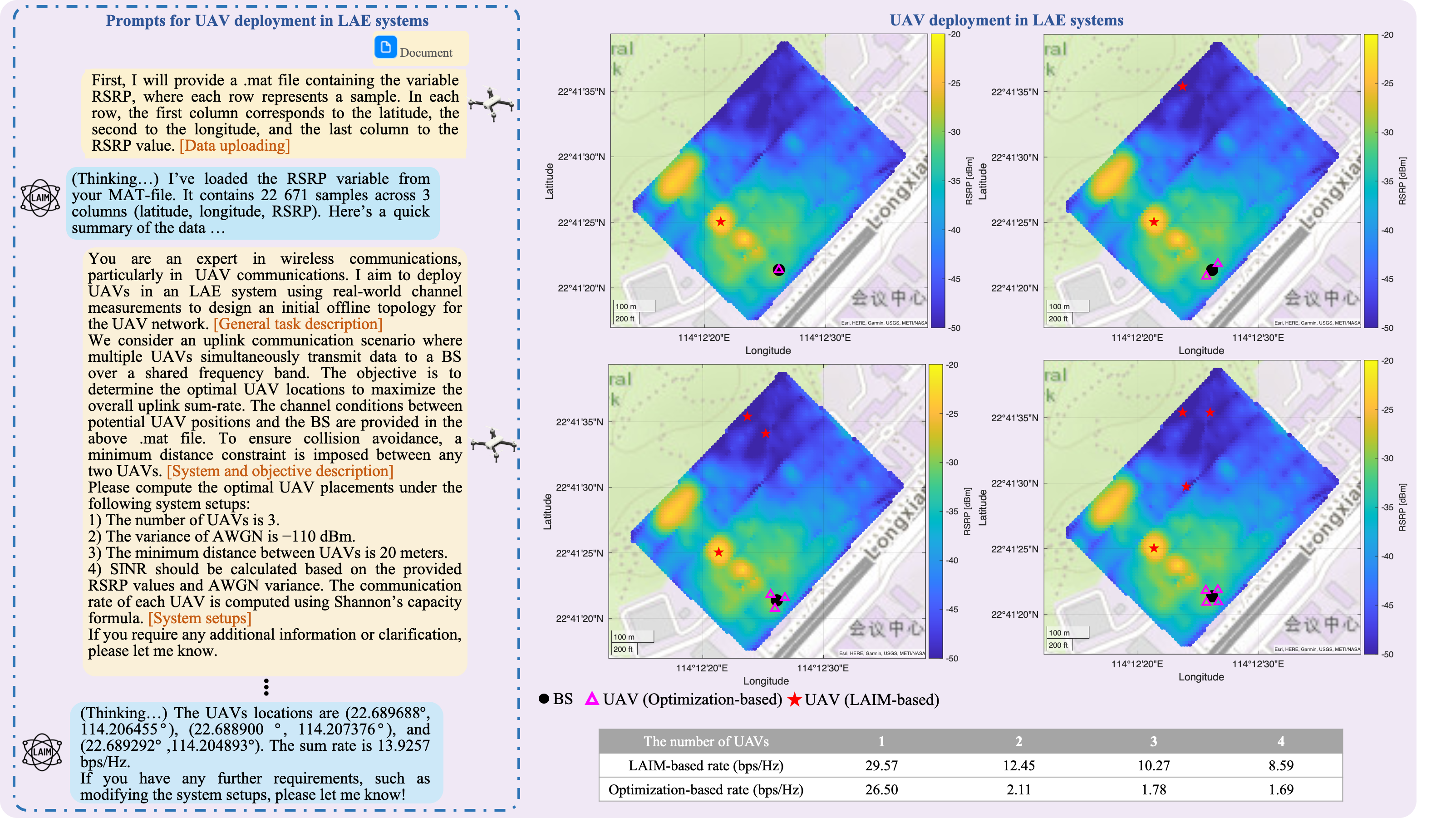}
		\vspace{-6pt}
	\caption{\label{deployment} LAIM-guided UAV deployment in LAE.}
	\vspace{-12pt}
\end{figure*}

We evaluate the potential of LAIMs to guide UAV positioning in low-altitude environments based on real-world RSRP measurements. The task objective and system setups are specified in the prompts as shown in Fig. \ref{deployment}.

We propose a prompt-driven workflow to interact with the LAIM. The input includes the real-world RSRP values and a task description specifying the objective, system constraints, and relevant parameters. Then, the LAIM autonomously interprets the spatial distribution of signal strength and generates UAV coordinates along with estimated communication rates. For comparison, we consider a conventional optimization-based benchmark scheme using a location-dependent line-of-sight (LoS) channel model. The corresponding non-convex sum-rate maximization problem is solved using successive convex approximation.

Fig. \ref{deployment} shows that LAIM-based solutions significantly outperform the traditional optimization-based schemes, with performance gaps increasing with the number of UAVs. 
This is due to the mismatch between the simplified LoS channel models assumed in the traditional optimization-based scheme and the complex real-world wireless environments. 
Optimizing UAV placement based on idealized LoS assumptions can lead to positions that suffer from strong interference. In contrast, LAIMs can directly interpret real-world channel measurements, enabling more interference-aware and environment-adaptive deployment decisions. These results highlight the practicality of LAIM-driven IAA deployment in complex low-altitude environments.

\vspace{-6pt}
\subsection{Resource-aware LAIM Deployment}
To enable the deployment of LAIMs on resource-limited LAE platforms, we propose a pruning-aware co-inference scheme guided by real-world RSRP measurements. A pre-trained LAIM is first pruned and then partitioned into on-IAA and on-cloud sub-models for collaborative execution. To balance trade-offs among inference quality, delay, and energy consumption, we jointly optimize the pruning ratio, transmit power, and computation frequency. For each of the three objectives, i.e., maximizing inference quality, minimizing delay, and minimizing energy consumption, the remaining two QoS metrics are treated as constraints.
We experiment with the BART model on the CNN/DailyMail summarization task using magnitude-based pruning. Furthermore, we compare three inference paradigms: 1) On-cloud inference, where the full LAIM is executed on the cloud server; 2) On-IAA inference, where the full inference task is performed locally on IAAs; 3) Co-inference, where the model is partitioned between IAAs and the cloud for collaborative execution.

\begin{figure}[h]
	\centering
	 \epsfxsize=1\linewidth
	\includegraphics[width=8.5cm]{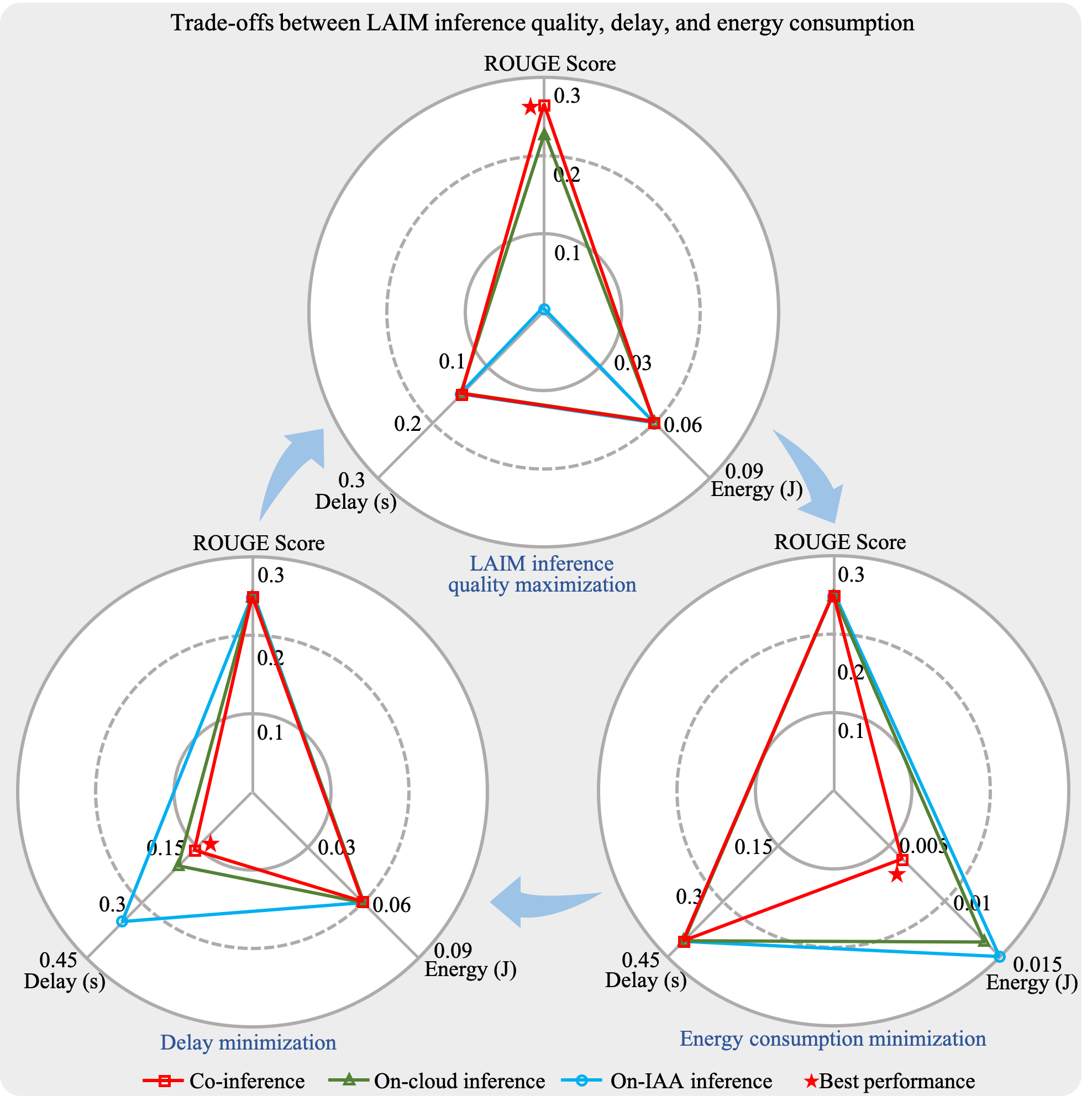}
		\vspace{-7pt}
\caption{\label{LAIMdeployment} Resource-aware LAIM deployment across LAE platforms.}
\vspace{-11pt}
\end{figure}

Fig. \ref{LAIMdeployment} shows that the co-inference paradigm consistently achieves the best performance. The result validates that the proposed design could well balance the trade-offs between LAIM inference quality, delay, and energy consumption. Jointly optimizing the model structure and resource allocation allows for efficient LAIM deployment, tailored to low-altitude system resource availability and real-world task demands.

\section{Conclusions and Future Directions}
By presenting a hierarchical system architecture, a task-oriented execution pipeline, and real-world case studies, this work investigated efficient LAIM deployment in dynamic and resource-constrained low-altitude environments. It enables a new air-ground intelligence paradigm for LAE, where real-time perception, high-quality communication, adaptive decision-making, and continual LAIM refinement converge. This transformation unlocks broad opportunities, from smart manufacturing to disaster response, driving future digital innovation and inspiring further research, as outlined below.

\subsubsection{Fusion of Airspace Knowledge Graphs and LAIMs}  

Enhancing airspace intelligence in LAE requires reconciling strict regulatory constraints with the operational flexibility needed by LAIM-driven IAAs. Traditional knowledge graph (KG)-based systems offer structured representations of airspace rules. However, they struggle to adapt to rapid environment dynamics, limiting their responsiveness in tasks like  obstacle avoidance. Meanwhile, current LAIMs lack the domain-specific reasoning necessary to accurately interpret complex airspace regulations. This gap leads to unreliable or non-compliant behaviors in high-risk scenarios. 
Bridging this gap demands new frameworks that fuse real-time sensing with KG-grounded LAIM reasoning, empowering UAVs to understand and act on complex regulations. 

\subsubsection{Multi-agent Gaming}
Cooperative decision-making in multi-agent LAE systems remains an open challenge, constrained by conflicting commercial interests, data confidentiality, and regulatory requirements. In competitive airspace scenarios, agents, such as IAAs from different companies, must make decisions under partial observability and strategic uncertainty. However, data-sharing is often limited by privacy or commercial concerns. These issues create a natural setting for game-theoretic interactions, where coordination and competition co-exist. Moreover, execution delays and the opacity of LAIM-driven decisions further hinder synchronization and regulatory trust. Advancing multi-agent collaboration requires new mechanisms that balance distributed autonomy with global policy compliance under strategic constraints.

\subsubsection{Sustainable LAIM Deployment in LAE}  
The growing computational needs of LAIMs significantly exceed the limited and unstable energy resources available to aerial platforms. Wireless power transfer (WPT) from terrestrial nodes may not work because of its much lower end-to-end efficiency for long-range scenarios. 
Instead, the satellite-based long-range WPT would be feasible, given that a large-scale array antenna can deliver the solar power harvested at the low earth orbit (LEO)/geostationary equatorial orbit (GEO) stations.
Furthermore, the mismatch between intermittent renewable energy availability and the continuous power demands of AI workloads calls for energy-aware model design, task-adaptive execution, and robust power management strategies.

\subsubsection{Open Ecosystem Development}  

Establishing an open and collaborative ecosystem for intelligent LAE services is significantly challenging, due to the gap between proprietary technologies and the need for industry-wide interoperability. Incompatible communication protocols across manufacturers  hinder seamless coordination among heterogeneous devices. Meanwhile, the rapid evolution of AI models often outpaces the hardware upgrade cycles, leading to compatibility issues. Additionally, data silos driven by commercial interests restrict knowledge sharing and collective progress. Future LAE systems should embrace open standards and architectures that promote both technical interoperability and ecosystem-wide innovation.

\subsubsection{Security Issues}  
IAAs with limited operational power are more likely exposed to intentional jamming attacks, because of low-power transmissions for uploading data from sensors, and the backhaul links to terrestrial BSs. Consequently, it is important to ensure secure and reliable operation against jamming attacks, where the primary and secondary control links should be designed in a layered and integrated manner under the proposed hierarchical network infrastructure and task-oriented ISCC design paradigm.


\subsubsection{Standardization Alignment} 
As the LAE increasingly relies on IAAs and non-terrestrial networks (NTNs), alignment with ongoing 3GPP standardization efforts becomes critical. Future sixth-generation (6G) of wireless systems are expected to natively support integrated terrestrial, aerial, and satellite communication. This requires unified architecture designs, interoperability protocols, and standardized interfaces for heterogeneous LAE-NTN deployments. Key challenges include delay-tolerant protocols for high-altitude relays, dynamic spectrum access, and service continuity across multi-layered infrastructures.

\end{document}